\documentclass[review]{elsarticle}

\usepackage{lineno,hyperref}
\modulolinenumbers[5]

\journal{Journal of Nuclear Instruments and Methods in
Physics Research A,}

\begin{document}

\begin{frontmatter}

\title{Study of the performance of a compact sandwich calorimeter for the instrumentation of the very forward region of a future linear collider detector\tnoteref{n1}}
\tnotetext[n1]{on behalf of \textbf{FCAL} Collaboration}

\author[iss]{V. Ghenescu\corref{cor1}}
\cortext[cor1]{Corresponding author}
\ead{veta.ghenescu@cern.ch}

\author[tau]{Y.Benhammou}

\address[iss]{Institute of Space Science,Bucharest-Magurele, ROMANIA}
\address[tau]{Tel Aviv University, TelAviv, ISRAEL}

\begin{abstract}
The \textbf{FCAL} collaboration is preparing large scale prototypes of special calorimeters to be used in the very forward region at a future linear electron positron collider for a precise and fast luminosity measurement and beam-tuning. These calorimeters are designed as sensor-tungsten calorimeters with very thin sensor planes to keep the Moliere radius small and dedicated FE electronics to match the timing and dynamic range requirements.
A partially instrumented prototype was investigated in the CERN PS T9 beam in 2014 and at the DESY-II Synchrotron in 2015. It was operated in a mixed particle beam (electrons, muons and hadrons) of 5 GeV from PS facilities and with secondary electrons of 5 GeV energy from DESY-II. The results demonstrated a very good performance of the full readout chain. The high statistics data were used to study the response to different particles, perform sensor alignment and measure the longitudinal shower development in the sandwich. In addition, Geant4 MC simulations were done, and compared to the data.

\end{abstract}


\end{frontmatter}

\linenumbers

\section{Introduction}

\begin{figure}[h!]
  \centering
   \includegraphics[width=0.55\textwidth]{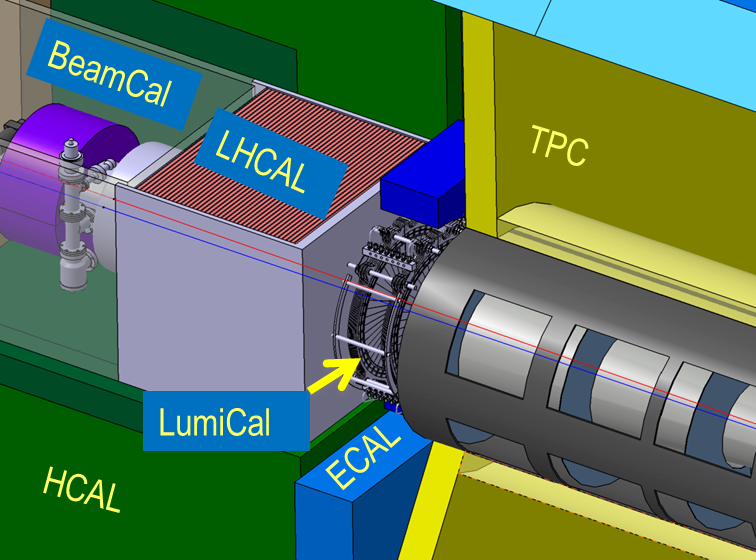}
\caption{The very forward region of the ILD detector.
LumiCal, BeamCal and LHCAL are carried by
the support tube for the final focusing quadrupole QD0 and the beam-pipe.
TPC denotes the central tracking chamber, ECAL the electromagnetic and
HCAL the hadron calorimeter. }
\label{fig:Forward_structure}
\end{figure}

In the future linear collider detectors the forward region is instrumented with two low angle electromagnetic calorimeters, denoted hereafter as \textbf{LumiCal} and \textbf{BeamCal}.  The Luminosity Calorimeter (\textbf{LumiCal}) precisely counts the number of Bhabha events and enables the measurement of the luminosity spectrum using the acollinearity angle of Bhabha scattering\cite{cdr_clic_1}. The Beam Calorimeter (\textbf{BeamCal}) extends the angular coverage of the forward calorimeters down to polar angles of about 10 mrad. Both calorimeters are centered around the axis of the outgoing beams and consist of 3.5 mm-thick tungsten plates, each corresponding to about one radiation length, interspersed with segmented sensor layers. All detectors in the very forward region have to tackle relatively high occupancy, requiring dedicated front-end electronics. The forward region layout foreseen for the ILD detector is shown in Figure~\ref{fig:Forward_structure}.
Prototype detector planes assembled with dedicated FE and ADC ASICs for \textbf{LumiCal} and
for  \textbf{BeamCal}  have  been  built.   In  this  paper,  results  of the \textbf{LumiCal}  performance  following  tests  in  a mixed particle and in an electron beam are reported.

\section{Testbeam instrumentation}
During 2014 and 2015, the \textbf{FCAL} collaboration has successfully commissioned and operated, in the T9 experimental area at the CERN PS facilities and in the 21 beam line area at the DESY II Synchrotron, the first \textbf{LumiCal} multi-layer  prototype. A multi-plane module with four sensor planes was investigated in a mixed particle and in an electron beam. Different detector configurations were used during the testbeams, with the active sensor layers always separated by one or two absorber layers.

\subsection{Testbeam at CERN}
\begin{figure}[h!]
  \centering
   \includegraphics[width=0.9\textwidth]{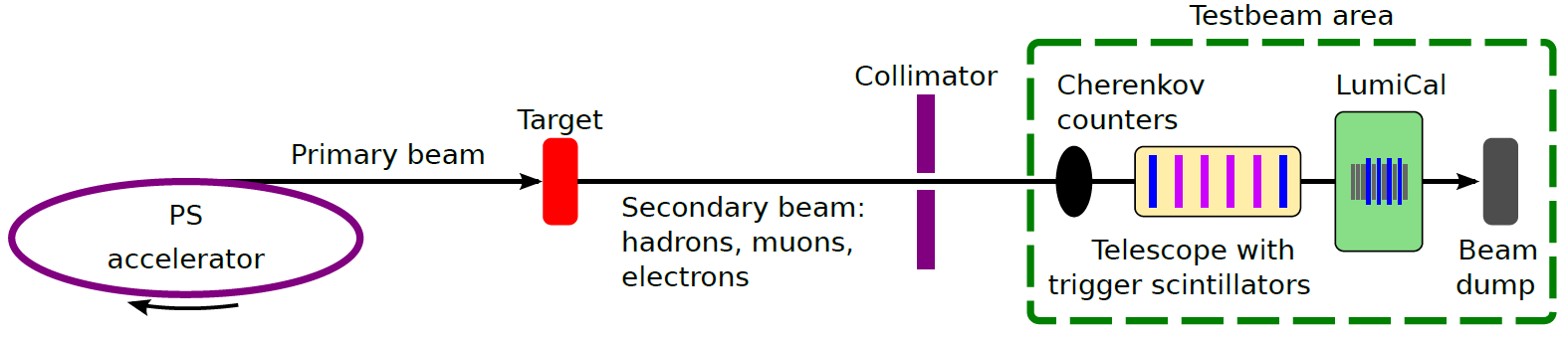}
\caption{Overall, simplified view (not in scale) of the testbeam facility. }
\label{fig:ps_line}
\end{figure}

The testbeam was performed in October 2014 at the CERN PS facilities using a secondary beam with muons, pions, hadrons and electrons with momentum of 5GeV/c. The beam momentum can be fixed by the magnetic field of the beam delivery electromagnets while the horizontal spatial spread is reduced by a set of collimators. The simplified overall (without magnets) of the PS facilities is presented in Figure~\ref{fig:ps_line}.

\begin{figure}[h!]
\begin{minipage}[t]{0.5\linewidth}
  \centering
   \includegraphics[width=1\columnwidth]{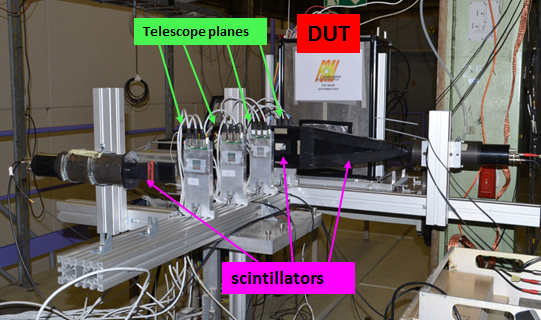}
\caption{Overall photograph of the testbeam area instrumentation. }
\label{fig:cern_set-up}
\end{minipage}
  \hspace*{0.01\linewidth}
  \begin{minipage}[t]{0.5\linewidth}
  \centering
   \includegraphics[width=0.42\columnwidth]{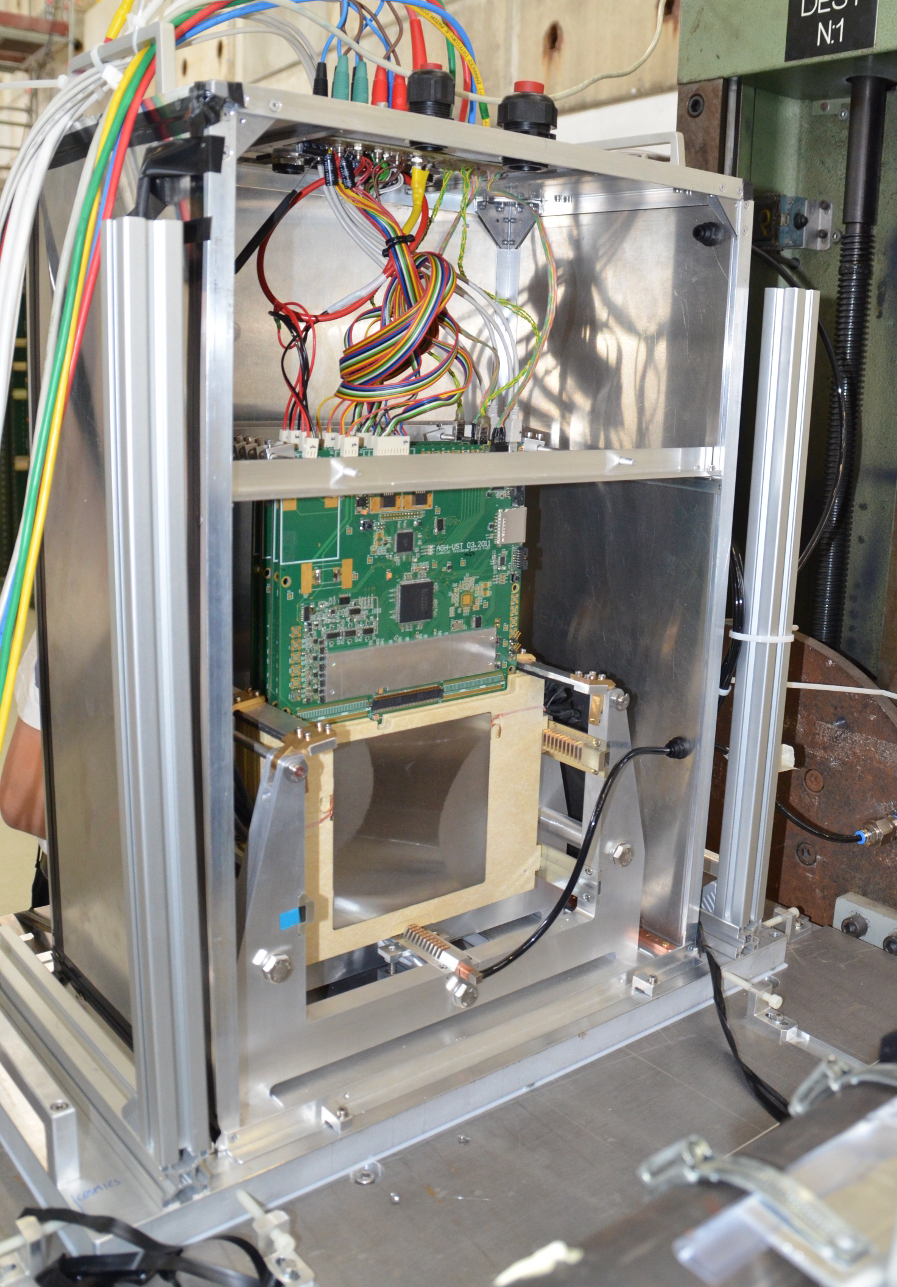}
\caption{Overview of the mechanical structure with the tungsten absorbers and the detector planes. }
\label{fig:dut_cern}
\end{minipage}
\end{figure}

\begin{figure}[h!]
  \centering
   \includegraphics[width=0.35\textwidth,angle=90]{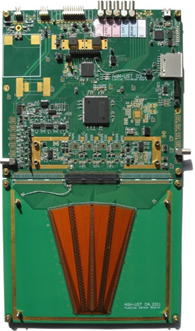}
\caption{Photograph  of  a  detector  plane  with  a  silicon  sensor  connected.   The  total  size  amounts  to 34$\times$21~cm$^2$.}
\label{fig:pcb_old}
\end{figure}

 In order to enable the particle track reconstruction, a multilayer tracking detector, so-called telescope, was developed by the Aarhus University. The telescope utilizes the Mimosa26 chips, a MAPS with fast binary readout \cite{mimosa}. The Mimosa26 utilizes a binary readout indicating the pixel signals exceeding the preset discrimination level. The pixel matrix is read continuously (without triggering) providing a complete frame every 115.2~$\mu$s. The data are gathered, triggered and stored by a custom DAQ system, based on the PXI crate, developed by the Aarhus University in collaboration with Strasbourg University. Four telescope planes, each comprising one Mimosa26 chip, were set upstream of the \textbf{DUT} as shown in Figure~\ref{fig:cern_set-up}. 
	The trigger system consists of three scintillators with compact photomultipliers. Two solid  5$\times$5mm$^2$ scintillators were placed upstream and downstream the telescope and one, with a 9~mm diameter circular hole, was placed just before
the last telescope plane. 
The \textbf{DUT} (LumiCal detector prototype), was prepared using four detector planes always separated by two absorber layers as shown in Figure~\ref{fig:dut_cern}. 
The complex mechanical structure developed by CERN allows to install the tungsten absorber or active sensor layer with a precision of a few tens of micrometers. A fully assembled detector plane is shown in Figure~\ref{fig:pcb_old} \cite{readout_board}. The prototype system consists of a silicon sensor covered by kapton fanout and front-end electronics.
Due to high energy depositions in the \textbf{LumiCal} calorimeter, standard 320~$\mu$m thick silicon sensors were chosen as a compromise between the capacitance and the amount of generated charge. \textbf{LumiCal} sensors have been designed at the Institute of Nuclear Physics PAN in Cracow \cite{silicon_design} and manufactured by Hamamatsu Photonics. The sensor is made from a N-type silicon, with p$^+$pads with a thin aluminum metalization. 
The sensor tile,  as the smallest sensor unit produced, includes 256 separate sensor pads. For this experiment 32 channels readout module was used.

\subsection{Testbeam at DESY}

\begin{figure}[h!]
\begin{minipage}[t]{0.5\linewidth}
  \centering
   \includegraphics[width=0.75\columnwidth]{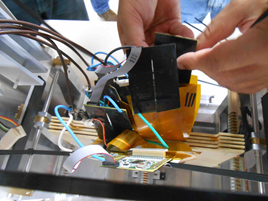}
\caption{The four sensors glued on the tungsten inside the permaglass frames. It is possible to see the orange read out kapton foils and the APV25 chips.}
\label{fig:desy_dut}
\end{minipage}
  \hspace*{0.01\linewidth}
  \begin{minipage}[t]{0.5\linewidth}
  \centering
   \includegraphics[width=1\columnwidth]{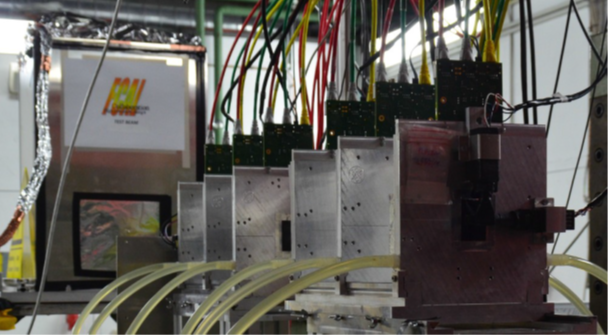}
\caption{The telescope composed of six silicon planes.}
\label{fig:desy_exp_set_up}
\end{minipage}
\end{figure}

During October 2015, four detectors were installed on the tungsten layers and installed inside the same frame as we used during the 2014 test beam as shown in Figure~\ref{fig:desy_dut}. The whole system was installed in 5 GeV electron beam at DESY. The beam line was composed of scintillators for the trigger and six planes of silicon used as a telescope \cite{eutelescope}  as shown in Figure~\ref{fig:desy_exp_set_up}.

\begin{figure}[h!]
  \centering
   \includegraphics[width=0.9\textwidth]{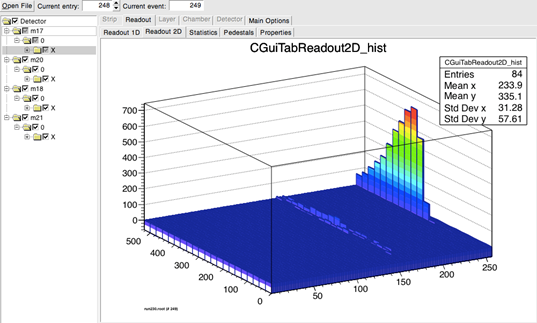}
\caption{ADC counts has a function of the position and the time. Here the channel 247 has been hit and the signal (after shaping) is starting at 300 ns.}
\label{fig:adc_256}
\end{figure}

A crucial point in the measurements was the synchronization between the data acquired with the detectors under investigation and that of the tracker. The latter was used to measure the local detection efficiency and evaluate the level of cross-talk between neighboring pads as function of the particle impact location. This was effectively achieved employing the RD51 Scalable Readout System (SRS). 
The data acquisition of both systems (silicon tracker and LumiCal detector) was triggered by the same scintillator signals and performed with a single SRS front-end card (FEC) \cite{RD51}. For each triggered event, the charge accumulated by all channels was stored. For each channel, the charge was sampled in 25 bins of 25 ns each. The mmDAQ1 online data acquisition software was used to store the synchronized data on a PC for further analysis. 
Using the online, it is possible to visualize events. On the Figure~\ref{fig:adc_256}, we can see a typical event: the ADC counts (pedestal subtracted) as a function of the channel number (between 1 and 256) and the time (between 0 and 500 ns). 

\begin{figure}[h!]
\begin{minipage}[t]{0.5\linewidth}
  \centering
   \includegraphics[width=0.9\columnwidth]{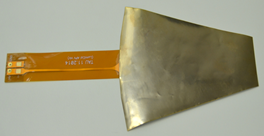}
\caption{The high voltage kapton foil and its trace to the connector. }
\label{fig:hv_fanout}
\end{minipage}
  \hspace*{0.01\linewidth}
  \begin{minipage}[t]{0.5\linewidth}
  \centering
   \includegraphics[width=1\columnwidth]{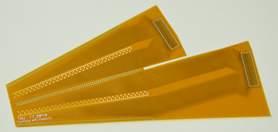}
\caption{The fan out kapton foil and the two 130 pin connectors. We printed two different path lengths to check the signal attenuation and cross talk over 5 and 10 cm.}
\label{fig:ch_fanout}
\end{minipage}
\end{figure}

The aim of the test beam was to fully equipped four silicon sensors and installed these sensors in between two tungsten planes, namely within one mm. We modified the read out kapton foil including two 130 pin connectors (Figure~\ref{fig:ch_fanout})  so we will be able to read all the pads of a sensors. The high voltage was originally injected from on the 3.5 mm PCB but to cope with the one mm, we designed a thin high voltage kapton foil as shown in the Figure~\ref{fig:hv_fanout}.
In order to mechanically support the sensor and the two kapton foils, we designed an envelope, which should also fit within 1 mm. We used two different techniques to produce the envelope: 3D printing and carbon fiber. It appears that the 3D printed envelop is not rigid enough so we finally choose the carbon fiber option, which allows thickness of ~200~$\mu$m and a good rigidity.

\begin{figure}[h!]
\begin{minipage}[t]{0.5\linewidth}
  \centering
   \includegraphics[width=1\columnwidth]{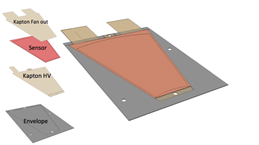}
\caption{A drawing of the different layer of the sub millimeter detector. }
\label{fig:desy_sensor}
\end{minipage}
  \hspace*{0.01\linewidth}
  \begin{minipage}[t]{0.5\linewidth}
  \centering
   \includegraphics[width=0.9\columnwidth]{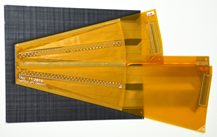}
\caption{The glued sensor with the two kapton foil and the carbon fiber envelope.}
\label{fig:desy_fibre}
\end{minipage}
\end{figure}
The whole detector (see Figure~\ref{fig:desy_sensor} including the carbon fiber, the two kapton foils and the sensor) thickness was measured in different positions around 900~$\mu$m.
An important effort has been realized on the gluing layers thickness and on the wire bonding loop size: some tools have been developed in order to be able to apply glue layers of 20 to 30 ~$\mu$m. We developed an expertise to maintain the wire bonding loop size around 50 ~$\mu$m.

In order to read all the pads of each sensor (128 channels) we used the APV25 chip. One hybrid contains 128 channels, which each are AC coupled to one read-out strip of the detector. Each channel contains a pre-amplifier, CR-RC shaper with adjustable shaping time and has a spark protection system implemented. The hybrid records the value of the shaped signal every 25 ns and stores it in an analog memory consisting of a 192 celled pipe-line. The signal is read out upon a trigger signal. The number of samples in one acquisition window is user defined, and the trigger time delay must hence be adjusted by the user to synchronize the trigger with the acquisition window. 
The APV hybrids are connected to the adapter boards using commercial HDMI cables. One cable can serve two APV hybrids (a master and a slave card) that can be connected through a 16-lead flat cable. The hybrid cards are connected to the detector ground through two low-ohmic low-profile RF coaxial connectors, which also serve to fix the hybrid card mechanically. 

\section{Results}
\begin{figure}[h!]
  \centering
   \includegraphics[width=0.75\textwidth]{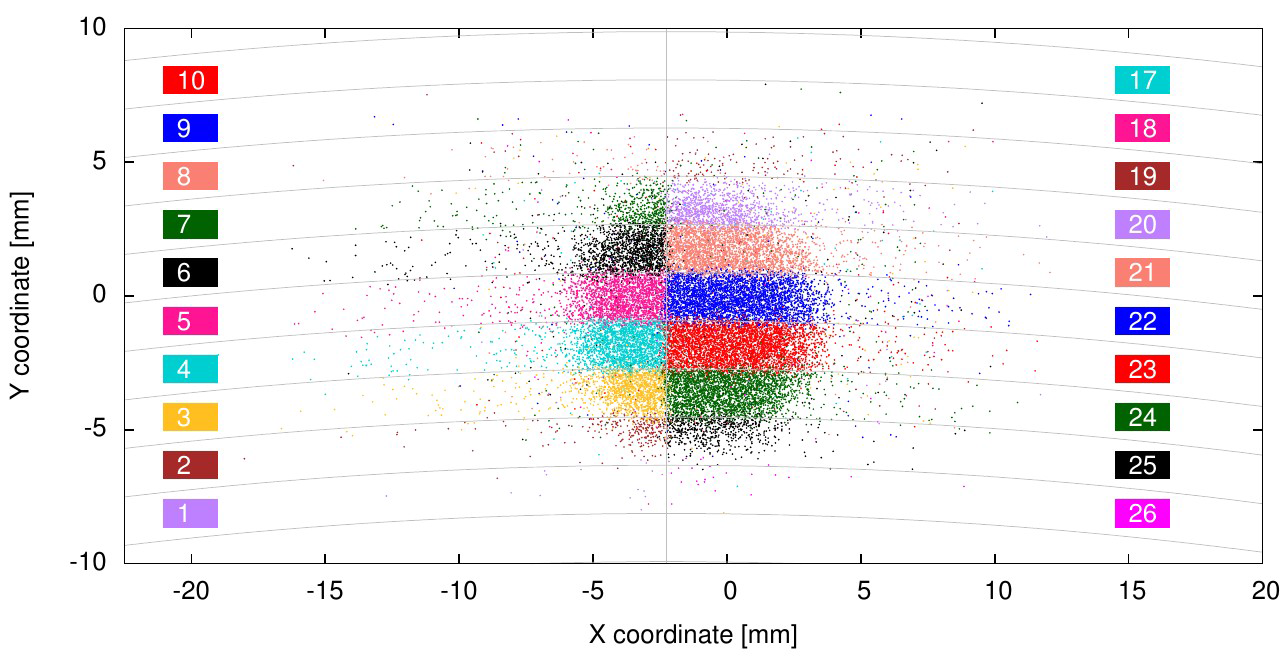}
\caption{Distribution of the tredicted impact points on pads with a color coded signal. }
\label{fig:impact_points}
\end{figure}

\begin{figure}[h!]
  \centering
   \includegraphics[width=0.75\textwidth]{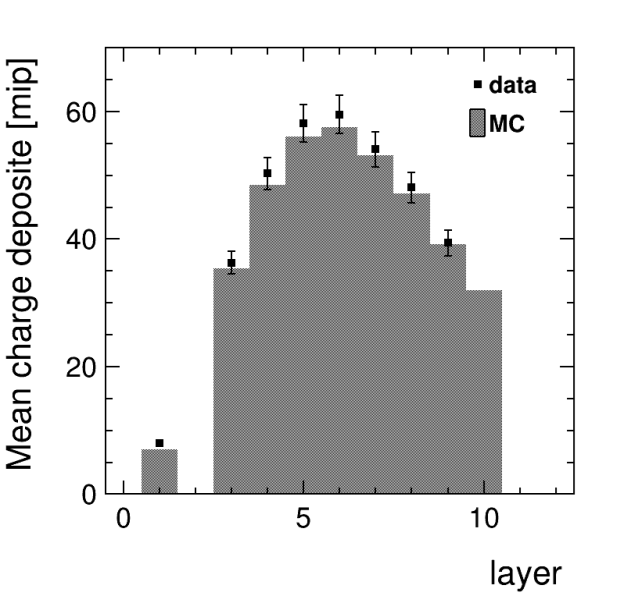}
\caption{Average energy deposited in the instrumented area of the LumiCal detector prototype as a function of the number of tungsten absorber layers. }
\label{fig:long_shower}
\end{figure}

The testbeam campaigns were done to investigate the performance of the calorimeter prototype. Data were taken for different pads and also for regions covering pad boundaries. The impact point on the sensor is reconstructed from the telescope information. Using a color code for the signals on the pads the structure of the sensor becomes nicely visible, as seen in Figure~\ref{fig:impact_points}.
During the  test beam, the response of the readout chain to electromagnetic shower was
studied. In order to analyze the electromagnetic shower development the energy deposited in each sensor plane from all configuration were obtained. The shower can be therefore sampled up to the 10$^{th}$ layer with the target spatial sampling resolution of one radiation length, by combining the data obtained in all three configurations. The measurement results from October 2014 test beam  were compared with prediction of GEANT4 Monte Carlo simulations where the experimental setup was implemented. A good agreement was found between them as shown in Figure~\ref{fig:long_shower}.

\section{Conclusions}
Data taken in tests beam were processed using various algorithms to investigate the response of large-area pad sensors assembled with a full readout chain. A first \textbf{LumiCal} multi-layer silicon-sensor prototype was tested at the PS CERN facilities with a mixed-particle beam and at the DESY II Synchrotron with a 5 GeV electron beam. The MC simulation of the testbeam was performed with a fair level of detail and the experimental data are in good agreement with MC simulations. The obtained results indicate an excellent working performance of all components (silicon sensors, front-end and readout electronics) and fulfilled the requirements for \textbf{LumiCal} calorimeter. Using new technology and expertise and each construction step, we succeeded to develop, produce and test detector less than 1 mm thick modules (900~$\mu$m). The first results of a four layers fully equipped have started and results are promising. 

\section*{Acknowledgments}
This work was partially supported by the Romania UEFISCDI agency under PN-II-PT-PCCA-2013-4-0967 project, the Israel Science Foundation (ISF), the German-Israel Foundation (GIF) and I-CORE.

\end{document}